\documentclass[aps,prl,twocolumn,nopacs,superscriptaddress,groupedaddress]{revtex4}

\usepackage{amssymb}
\usepackage{color,graphicx}
\usepackage{amsmath}
\usepackage{amsbsy}
\usepackage{amsthm}
\usepackage{bbm}
\usepackage{bm}
\usepackage{epsfig}
\usepackage{lscape}
\usepackage{float}
\usepackage{subfigure}

\begin{document}

\title{Ferromagnetic detection of moduli dark matter}

\author{A. Vinante}
\email{anvinante@fbk.eu}
\affiliation{Istituto di Fotonica e Nanotecnologie, CNR - Fondazione Bruno Kessler, I-38123 Povo, Trento, Italy}

\date{\today}

\begin{abstract}
We propose a scheme to detect light scalar moduli dark matter, based on measuring the change of magnetization induced in a macroscopic hard ferromagnet.
Our method can probe moduli dark matter at the natural coupling to the electron mass over several orders of magnitude in the moduli mass. The most attracting feature of the proposed approach, compared to mechanical ones, is that it relies on a nonresonant detection, allowing to probe a much wider region of the parameter space. This is a crucial point, as long as the theory is not able to predict the moduli mass.
\end{abstract}

\maketitle

\textit{Introduction -} String theory predicts the possible existence of light scalar fields called moduli. Under the assumption of supersymmetry breaking, moduli acquire a mass, which is model-dependent and may lie in a large range spanning more than $20$ orders of magnitude \cite{dimopoulos}. An experimental signature of moduli would be the oscillation of fundamental constants coupled to the moduli field. Relevant cases are the coupling of moduli to electrons and to photons, which lead to an oscillation of the electron mass $m_e$ and the fine structure constant $\alpha$, respectively \cite{arvanitaki}. 

Very light scalars such moduli constitutes a good candidate for cosmological dark matter, provided that the particle mass is at least $10^{-22}$ eV \cite{gruzinov}. Assuming that moduli have a well-defined mass $m_\phi$ and that they make up the totality of local dark matter, they will produce an oscillation of $m_e$ or $\alpha$ centered around the rest mass frequency $f_\phi=m_\phi/2\pi$ with a fractional width $\Delta f_{\phi}=10^{-6} f_\phi$, due to the velocity spread of dark matter. The amplitude of the fractional change $h$ is given, (in natural units), by:
\begin{equation}
h = d_{i} \frac{{\sqrt {4\pi G\rho _{DM} } }}{{m_\phi  }}
\end{equation}
where $\rho_{DM}=0.3$~GeV/cm$^3$ is the local dark matter density, $G$ is the gravitational constant and $d_i$ is the relevant coupling, with $i=m_e$ or $i=\alpha$ respectively. A natural parameter space can be defined assuming a supersymmetry breaking at 10 TeV. For the coupling to the electron mass this corresponds to $d_i = d_{m_e} \simeq  10^{-5}  f_\phi /$Hz, while the natural $d_\alpha$ is 7 orders of magnitude smaller. At the natural coupling level, the fractional change of $m_e$ is $h\sim 10^{-21}$, which makes the detection of hypothetical moduli dark matter roughly as challenging as the one of detecting gravitational waves \cite{GW}.

Detection of moduli dark matter \cite{arvanitaki} through the mechanical effect exerted on a macroscopic mechanical system has been recently discussed. First relevant upper limits have been established by the AURIGA gravitational wave detector, probing for the first time the natural space for $d_{m_e}$ in a narrow band of $f_\phi$ around 900~Hz \cite{branca}. The detection concept is that moduli field, by modifying the fundamental constants $m_e$ and $\alpha$, will also change the size of atoms, as Bohr radius $a_0 \sim 1/(\alpha m_e)$. Extension to macroscopic solid objects with length $L$ leads to $\delta L/L \sim -\delta m_e/m_e - \delta \alpha/\alpha$.

Here, we propose to exploit the analog effect in a macroscopic ferromagnet. In this case the signal will arise from the variation of the elementary magnetic moment, which we will assume equal to $g \mu_B$, where $g$ is a moduli-independent Lande factor and $\mu_B=e \hbar/(2 m_e )$ the Bohr magneton. As $\mu_B \propto  1/m_e$, one expects a moduli-induced variation of the magnetization of a macroscopic sample $\delta  M/M \sim -\delta m_e/m_e$. The effect has thus the same absolute magnitude, in terms of fractional change $h$, as in the mechanical case. However, in contrast with the latter, the dynamics of internal magnetic modes in a ferromagnet is very fast, which naturally suggest a wideband nonresonant strategy to detect the moduli signal. This is the main point of our proposal. While we find that the absolute sensitivity would be locally worse than using resonant mechanical techniques, our approach allows the exploration of a much larger range of the moduli mass.

\begin{figure}[!ht]
\includegraphics[width=8.6cm]{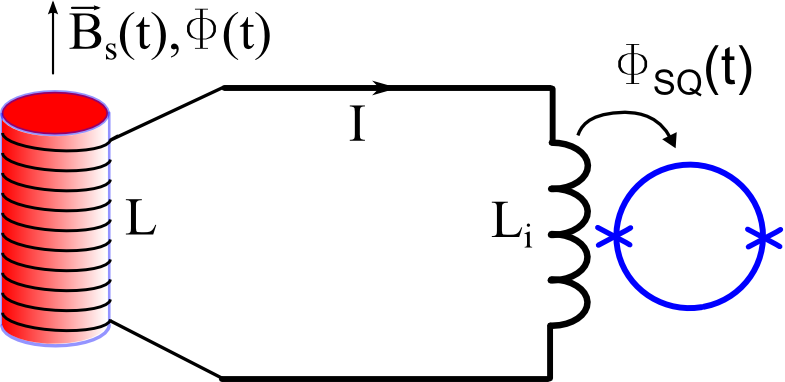}
\caption{Detection scheme. The moduli field induces a flux change $\Phi$ across the magnet, which is nonresonantly detected by a SQUID through a superconducting flux transformer.}  \label{figure1}
\end{figure}

\textit{Experimental proposal -} The scheme we propose is extremely simple. The internal magnetic field $B_S$ of a macroscopic hard ferromagnetic sample, of volume $V$, is probed by a superconducting pick-up coil connected to a SQUID magnetic flux detector (Figure 1).

To simplify the calculation and neglect border effects and demagnetization fields, we assume that the sample is a long and thin cylinder fully magnetized with the easy axis along the cylinder axis. The pick-up coil is a solenoid uniformly wounded around the magnet. This approximation can be practically obtained by using an array of long and thin rods, with pick-up coils connected in parallel. Under this assumption the internal field $B_S$ is equal to $\mu_0 M_S$, where $M_S$ is the static magnetization. 
The magnetic flux $\Phi$ threading the solenoid pick-up coil is related to the magnetization field $B_S$ by the following energy relation:
\begin{equation}
\frac{{\Phi ^2 }}{{2L}} = \frac{1}{{2\mu _0 }}\int {B_S ^2 } dV = \frac{{B_S ^2 V}}{{2\mu _0 }}   \label{flux1}
\end{equation}
where $L$ is the pick-up coil inductance and $V$ is the volume of the magnet. This equation is strictly valid when $B_S$ is replaced by the field $B(I)$ produced by a current $I$ in the pick-up coil. However, because of our assumptions, the magnetization field has the same uniform shape as $B(I)$, leading to the same relation between field and flux.
By letting the magnetization field $B_S$ vary as $B_S(t)=B_{S0}(1+h(t))$, where $h(t)$ is the fractional change induced by the moduli field, and combining this relation with Eq. (\ref{flux1}), we arrive at the flux signal induced in the pick-up coil:
\begin{equation}
\Phi \left( t \right) = h\left( t \right)B_{S0} \sqrt {\frac{{VL}}{{\mu _0 }}}.   \label{flux2}
\end{equation}
We neglect for the moment the fact that the volume of the magnet will also oscillate because of the moduli field. We will discuss this issue later.

The signal is then detected by a dc SQUID. We model the SQUID as a current amplifier, with input inductance $L_i$. We assume that the equivalent inductance $L$ of the pick-up coil is equal to the input inductance $L_i$, which constitutes the impedance matching condition. The SQUID noise can be expressed as an input current noise, with white spectral density $S_I$ \cite{SQUID}. SQUID back-action will cause an additional white noise contribution which we incorporate in the total noise. It is common to express the total SQUID noise as an equivalent energy resolution $\varepsilon = L_i S_I/2$. The best SQUIDs available today with large input coil, suitable for this experiment, are nearly quantum limited. For available SQUIDs the energy resolution is $\varepsilon  = N_{\hbar} \hbar$, with $N_{\hbar}\sim 10$ \cite{falferiSQUID}, but $N_{\hbar}\sim 1$ is in principle achievable. 

Under a nonresonant detection scheme one expects that the dominant source of noise is the SQUID itself. Therefore, we neglect other sources of noise at first order, but we will discuss this issue more in detail later. 
At the impedance matching condition, the current signal coupled into the SQUID input coil will be:
\begin{equation}
I\left( t \right) = \frac{{\Phi \left( t \right)}}{{2L}} = \frac{1}{2}h\left( t \right)B_{S0} \sqrt {\frac{V}{{L\mu _0 }}}    \label{Isignal}
\end{equation}
In the signal bandwidth $\Delta f_{\phi}$ the signal to noise ratio is given by:
\begin{equation}
SNR^2  = \frac{{I^2 }}{{S_I }} = \frac{{h^2 B_{S0} ^2 V}}{{8\mu _0 \hbar N_\hbar  \Delta f_{\phi}}}
\end{equation}
Note that the signal to noise ratio depends only on the static magnetization, the volume and the SQUID energy resolution. The actual value of the pick-up coil inductance $L$ has disappeared. 
The minimum detectable signal at SNR=1, within an integration time $\sim 1/ \Delta f_\phi$ will thus be:
\begin{equation}
h_{\min }  = \sqrt {\frac{{8\mu _0 \hbar N_\hbar  \Delta f_\phi}}{{B_S ^2 V}}} \label{hmin}
\end{equation}

\begin{table}
\begin{center}
\begin{tabular}{|c || c | c | c | c |}
\hline
Configuration &  $V (\mathrm{m}^3)$  & $B_S (\mathrm{T})$ & $N_\hbar (\hbar)$ & $t(\mathrm{yr})$ \\
\hline 
1  & $10^{-2}$ & 0.4  & 10 & 1  \\
\hline
2 & 1 & 1.3  & 1 & 1 \\
\hline
\end{tabular}
\end{center}
\caption{Parameters of the two experimental configurations considered in Fig. 2.} \label{tabella}
\end{table}

We consider two possible configurations, with main parameters listed in Table \ref{tabella}. Configuration 1 is a relatively simple small-scale experiment employing a hard ferrite with $B_S=0.4$~T and readily available SQUIDs. Configuration 2 is an advanced scaled-up version with $B_S=1.3$~T, corresponding to the magnetization of rare earth magnets and a quantum limited SQUID.

For the two configurations we calculate the minimum detectable signal after 1 year integration time, which is derived by Eq. (\ref{hmin}) by taking into account that the SNR referred to $h$ scales with $(t \Delta f_\phi)^{\frac{1}{4}}$ \cite{branca}.
The data are plotted in Figure 2, together with the natural limit for the $d_{m_e}$ coupling, and other relevant bounds from other experiments.

The configuration 1 will not be able to probe the natural region for the $d_{m_e}$ coupling. However, it will improve over the best wideband upper limit on moduli which can be inferred by fifth force experiments for frequencies below 1 MHz \cite{arvanitaki, adelberger}. 
In contrast, the configuration 2 will be able to probe the natural region for frequencies below a few MHz.
Notice that resonant mechanical experiments are able to set stronger upper bounds but only in a narrow bandwidth, unless very non straightforward schemes are implemented \cite{arvanitaki}. For instance, as shown in Figure 2, the AURIGA experiment has recently set a bound $\sim 3$ orders of magnitude stronger than the nominal natural limit in a bandwidth of roughly 100 Hz around 900 Hz \cite {branca}.
The key advantage of our strategy is clearly the wide bandwidth, which will allow to explore a much larger region of the parameter space.

\begin{figure}[!ht]
\includegraphics[width=8.6cm]{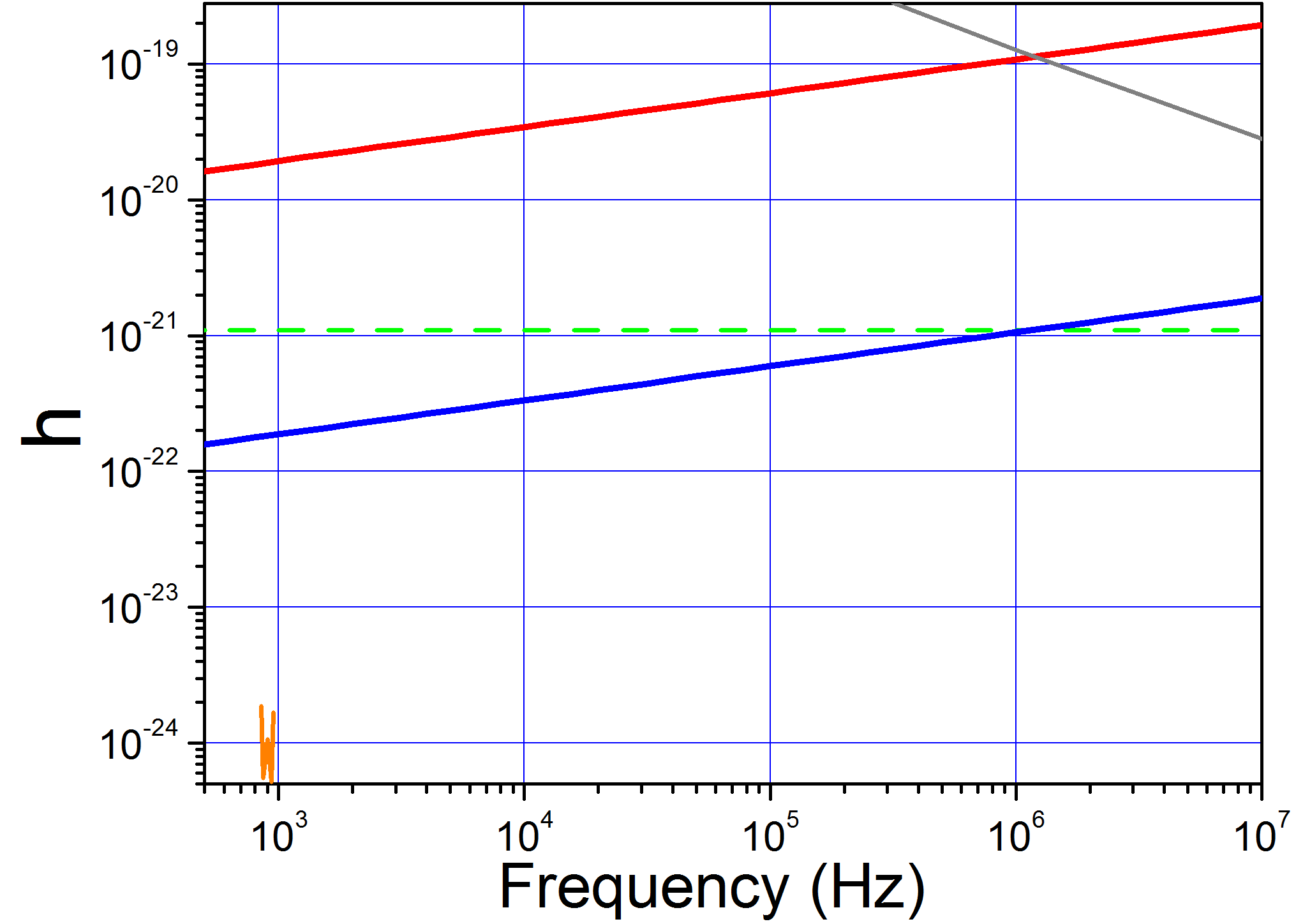}
\caption{Upper limit on the moduli-induced fractional change of the electron mass $h$, as a function of frequency, in the configurations 1 (top red line) and 2 (bottom blue line) after 1 year integration time. The green dashed line defines the natural parameter space for the $d_{m_e}$ coupling. The thin gray line represents the current upper bound from fifth force experiments \cite{arvanitaki, adelberger}. The thin orange line around 900~Hz represents the strong but narrowband limit recently set by the AURIGA detector \cite{branca}. }  \label{figure2}
\end{figure}
 
\textit{Possible signal cancellations - } Next, we discuss possible issues which could affect the simplified treatment described above. First we consider the signal, and its possible cancellation due to the moduli field affecting more quantities at the same time. We consider two effects. The first is that the SQUID actually measures a flux signal $\Phi_{SQ}=M_{SQ} I$ in natural units of flux quanta $\Phi_0=\hbar/2e$. Here, $M_{SQ}$ is the mutual inductance between the SQUID and the input coil. Therefore, the measured signal must be written as $\Phi_{SQ}/\Phi_0$, which turns out proportional to $e^2/m_e\sim \alpha^2/m_e$. This means that the detector is also sensitive to moduli coupled to $\alpha$ in addition to moduli coupled to $m_e$. The second effect is connected with the mechanical response of the system. At frequency much lower than the lowest mechanical resonant frequency, all characteristic lengths of a solid body are expected to oscillate coherently with the moduli field. Let us assume that all lengths just follow the Bohr radius, as assumed in Ref.~\cite{arvanitaki}. To see what happens we have to rewrite the measured signal in Eq.~(\ref{Isignal}) in a more explicit way, by writing $B_S=M_S/\mu0=N_S \mu_B/\mu_0 V$, with $N_S$ the number of spins. We have then:
\begin{equation}
\frac{{\Phi  }}{{\Phi _0 }} = \mu _0 \frac{{N_S \mu _B }}{V}\sqrt {\frac{{VL}}{{\mu _0 }}} \frac{{M_{SQ} }}{{2L}}\frac{1}{{\Phi _0 }}.
\end{equation}
Now, taking into account that any inductance (or mutual inductance) can be always written as $\mu_0$ times a characteristic length, we can conclude that:
\begin{equation}
\frac{\Phi}{{\Phi _0 }} \propto \frac{{\mu _B }}{{\Phi _0 a_0 }} \propto \alpha ^2 m_e ^0 
\end{equation}
In the last relation we have taken into account that $\mu_B/\Phi_0 \sim \alpha/m_e$ and $a_0\sim 1/(\alpha m_e)$
Thus, we arrive at the surprising result that the dependence on $m_e$ disappears, while the one on $\alpha$ doubles. However, we have to remark that the cancellation is effective only well below the mechanical resonances. For frequencies higher than the lowest longitudinal resonance the mechanical contribution will be almost completely suppressed, and the pure magnetic contribution from $\mu_B$, expressed by Eq. (\ref{Isignal}) will dominate. In conclusion, even if a complete cancellation may take place at low frequency, realistic experiments can be still proposed with a sensitive frequency range from $\sim  1$ kHz up to tens of MHz, where SQUIDs can be readily employed.

\textit{Other noise sources -} Another important issue is represented by additional noise contributions. We briefly discuss here four main noise sources: thermal magnetization fluctuations, thermal electrical noise from the detection circuit, eddy currents in the magnet and vibrational noise. 

Magnetization fluctuations in a saturated hard ferromagnetic material are intrinsically small, in sharp contrast with soft ferromagnetic materials employed to amplify magnetic fields. Hard ferromagnetic materials are characterized by a strong magnetic anisotropy, typically expressed by a anisotropy field $B_A$, which is associated to a ferromagnetic resonance frequency $\omega_{\mathrm{fm}}=\gamma B_A$ where $\gamma$ is the gyromagnetic ratio. Spin waves with frequency lower than the ferromagnetic resonance are forbidden by an energy gap. The thermal magnetization noise can be estimated by using the fluctuation-dissipation theorem combined with the phenomenological Landau-Lifshitz-Gilbert equation \cite{rugar}. For a single domain with volume $V$ fully magnetized along $z$, it is found that only tranverse magnetization fluctuations are relevant with spectral density given by:
\begin{equation}
S_{M_t }  = \frac{{4k_B T \alpha_d M_S }}{{\omega _{\mathrm{fm}} B_A V}}
\end{equation}
where $\alpha_d$ is a phenomenological damping parameter usually in the order of $0.01$.
In our detection scheme, transverse fluctuations give in principle no contribution to the measured signal. However, even assuming a worst case hypothesis that the effective fluctuations along the measurement axis $z$ are of the same order of magnitude of the transverse ones, and using typical values $\omega_{\mathrm{fm}}/2\pi= 700$~GHz, $\mu_0 M_S=1$~T, and setting $T=30$~K corresponding to the classical to quantum transition at $\omega=\omega_\mathrm{fm}$, one finds that the magnetization noise is less than $10^{-3}$ times the SQUID noise at the quantum limit. Of course, this estimation is very indicative, as it is based on the oversimplified assumption of single domain. A more reliable estimation of the actual amount of magnetic noise can be done by direct experimental measurements on small macroscopic samples fully magnetized.

Thermal noise can also arise by any additional dissipation in the detection circuit. For instance, dielectric or paramagnetic dissipation in the stray capacitance of the superconducting wire insulator leads to significant contribution to the total loss of high $Q$ LC resonators at audio frequency, with equivalent $Q$ factor of the order of $10^6$ at $T\sim 0.1$~K and $\omega/2\pi \sim 1-10$~kHz \cite{vinantethermal}. We can again use the fluctuation dissipation theorem by considering a resistance $\omega L/Q$ in the readout circuit. For $T=0.1$~K this leads to a contribution which is less than 10\% of the SQUID noise at the quantum limit.

Eddy current losses in the ferromagnet represent a huge contribution for metallic materials at low frequency. For a representative cylinder with radius $R\sim 1$~cm and electrical conductivity $\sigma\sim 10^6$ $\Omega$ (typical of neodymium-based magnets), the frequency at which skin effect becomes significant is $f_{c}= 1/(\pi \sigma \mu_0 R^2) \sim 3$~kHz. This means that, in contrast with magnetization fluctuations, the thermal energy $k_B T$ is now concentrated in a relatively narrow bandwidth $\sim $ kHz at low frequency leading to a huge spectral density. On the other hand, surface eddy currents will completely shield any internal field change, included the moduli signal, at higher frequency. To make a numerical estimation of the eddy current noise, one can again use the fluctuation-dissipation theorem. For a long and thin cylinder of radius $R$ and conductivity $\sigma$, elementary calculations show that eddy currents induced by a current in the coil are equivalent to an effective resistance:
\begin{equation}
r (\omega) = \frac {1}{8} \sigma \omega^2 \mu_0 R^2 L
\end{equation}
in series with the inductance $L$. At low frequency this leads to a current noise contribution in the SQUID input coil:
\begin{equation}
S_I= \frac{k_B T \sigma \mu_0 R^2}{8 L}.
\end{equation}
With $R\sim 1$~cm, $\sigma\sim 10^6$~$\Omega$ and $T=0.1$~K, the low frequency noise below the skin effect cutoff $f_{c}$ is $10^5$ times larger than the SQUID noise at the quantum limit.
Therefore, the experiment strictly requires insulating ferromagnetic materials.
This is a significant constraint, as the strongest rare-earth magnetic materials (like NdFeB or SaCo) are electrically conducting. On the other hand, insulating ferromagnets such as ceramic ferrites are usually weaker ($B_S\sim 0.4$~T). In addition, they typically lose their magnetic properties at low temperature, although hard ferrites exist which maintain their magnetization even at low temperature \cite{falferi}.

Finally, let us consider the ambient vibrational noise. Given the signal to be detected in the order of $10^{-21}$, the level of mechanical filtering required is comparable to that already achieved in gravitational wave detectors. While some care is needed to achieve a sufficient attenuation over a wide bandwidth with $f>1$ kHz, this should not represent a major obstacle for the experiment.

\textit{Alternative resonant LC scheme -} We have proposed a nonresonant scheme, as it marks the most relevant advantage of the magnetic approach over the mechanical one. However, the detection scheme can be easily converted to a resonant one, by introducing a capacitor in the circuit. 
In terms of sensitivity, a resonant scheme may provide some slight improvement over the nonresonant one. The SQUID additive wideband noise will be made negligible around the LC resonance, but the electrical thermal noise in the LC and the SQUID back-action noise will be still present. The foreseen improvement is at most one order of magnitude compared to the nonresonant scheme. 

In turn, a resonant ferromagnetic experiment will be roughly equivalent to a resonant mechanical experiment.
However, if the capacitor can be tuned in-situ, one may implement a scanning strategy, as proposed in \cite{arvanitaki} for a mechanical system. Compared to the mechanical case, the tuning can be done without bandwidth limitations and without changing the temperature, therefore allowing to maintain optimal noise properties. 

\textit{Conclusion -} We have shown that a simple experimental setup with a SQUID coupled to a macroscopic sample of hard ferromagnet under saturation is in principle an attracting system to detect moduli dark matter over a wide range of frequency from 1 kHz up to several MHz. In contrast with mechanical techniques, our method is intrinsically wideband, due to the very fast internal dynamics of hard ferromagnets. Among the open issues, it remains to assess whether a suitable ferromagnetic material can be found with the optimal properties. In summary, the proper ferromagnetic material must be electrically insulating, hard (high anisotropy), featuring the highest possible $M_S$ at cryogenic temperature. Finally, if large volumes are needed, practical aspects such as the availability of large samples and the cost need to be taken in to account.

\textit{Acknowledgments -} The author thanks Asimina Arvanitaki, Massimo Cerdonio and Paolo Falferi for useful discussions.


\begin{thebibliography}{<99>}

\bibitem{dimopoulos} S. Dimopoulos and G. F. Giudice, Phys. Lett. B379, 105 (1996), arXiv:hep-ph/9602350 [hep-ph].

\bibitem{arvanitaki} A. Arvanitaki, S. Dimopoulos, K.V. Tilburg, Phys. Rev. Lett. 116, 031102 (2016).

\bibitem{gruzinov} W. Hu, R. Barkana, and A. Gruzinov, Phys. Rev. Lett. 85, 1158 (2000), arXiv:astro-ph/0003365 [astro-ph].

\bibitem{GW} B.P. Abbott et al (LIGO Scientific Collaboration and Virgo Collaboration), Phys. Rev. Lett. 116, 061102 (2016).

\bibitem{branca} A. Branca, M. Bonaldi, M. Cerdonio, L. Conti, P. Falferi, F. Marin, R. Mezzena, A. Ortolan, G.A. Prodi, L. Taffarello, G. Vedovato, A. Vinante, S. Vitale, J.-P. Zendri, arXiv:1607.07327.

\bibitem{SQUID} P. Falferi, M. Cerdonio, M. Bonaldi, M. Mueck, A. Vinante, R. Mezzena, G.A. Prodi, and S. Vitale, J. Low Temp. Phys. 123, 275 (2001).

\bibitem{falferiSQUID} P. Falferi, M. Cerdonio, M. Bonaldi, R. Mezzena, G.A. Prodi, A. Vinante and S. Vitale, Appl. Phys. Lett. 93, 172506 (2008).

\bibitem{adelberger} E. Adelberger, B.R. Heckel, and A. Nelson, Ann. Rev. Nucl. Part. Sci. 53, 77 (2003), arXiv:hep-ph/0307284 [hep-ph].

\bibitem{rugar} B.C. Stipe, H.J. Mamin, C.S. Yannoni, T.D. Stowe, T.W. Kenny, and D. Rugar, Phys. Rev. Lett. 27, 277602 (2001).

\bibitem{vinantethermal} A. Vinante, R. Mezzena, G.A. Prodi, S. Vitale, M. Cerdonio, M. Bonaldi and P. Falferi, Rev. Sci. Instrum. 76, 074501 (2005).

\bibitem{falferi} P. Falferi, private communication.


\end{thebibliography}
\end{document}